\documentclass[12pt]{article}
\usepackage{epsf}
\usepackage[dvips]{graphicx}
\usepackage{subfigure}
\usepackage{bibspacing}
\renewcommand{\refname}{\normalsize REFERENCES}
\begin{document}
\vspace*{1cm}
\begin{center} 
To be published in
\end{center}
\begin{center} 
{\bf
A.~V.~Narlikar (Ed.) \\
``Frontiers in Superconducting Materials'' \\
Springer Verlag, Berlin, 2005 }
\end{center}
\vspace*{2.5cm}

\begin{center} 
{\bf \large SUPERCONDUCTIVITY OF MAGNESIUM DIBORIDE: THEORETICAL ASPECTS}
\vspace{2ex}

Thomas Dahm
\vspace{1ex}

Institut f\"ur Theoretische Physik, Universit\"at T\"ubingen, 
Auf der Morgenstelle 14,\\ D-72076 T\"ubingen, Germany

\end{center}
\vspace{3ex}

\noindent
{\bf 1. INTRODUCTION}
\vspace{1ex}

Our theoretical understanding of superconductivity in magnesium diboride 
(MgB$_2$) has made rapid progress since its discovery by Nagamatsu et 
al. \cite{Akimitsu}. Unlike superconductivity in the high-$T_c$ cuprates we are 
in a possession of a very clear picture of its superconducting state now
\cite{Canfield}. It 
seems clear that superconductivity is driven by electron-phonon interaction 
in MgB$_2$. More excitingly, it appears well established both theoretically 
and experimentally that the rare form of two gap superconductivity is realized 
in this compound. Two superconducting gaps of distinctively different size 
appear to exist on different disconnected parts of its Fermi surface. Since 
MgB$_2$ is the clearest example of two gap superconductivity to date, it 
makes it an interesting object for study and exploration. In the present 
work we want to review our present understanding of two gap superconductivity 
in MgB$_2$ from the theoretical perspective and discuss some of its 
consequences. The presence of these two different gaps gives rise to a 
number of anomalies and some of them shall be discussed here. 
Since the literature on superconductivity in MgB$_2$ has grown
rapidly, no complete coverage of experimental and theoretical work can
be made here. Instead, the theoretical results most crucial to our
understanding in the authors view will be discussed and a selection
of peculiar consequences will be presented.

Since much of our present understanding stems from band structure calculations 
and solutions of Migdal-Eliashberg equations, we are reviewing these results 
in the next section. The emergence and stability of two gap superconductivity 
in MgB$_2$ shall be discussed in section 3. In section 4 we will discuss 
consequences of this band structure picture for the upper critical field. 
Section 5 is finally devoted to the microwave conductivity. 
\vspace{2ex}

\noindent
{\bf 2. THE PICTURE SUGGESTED BY BAND STRUCTURE CALCULATIONS}
\vspace{1ex}

Magnesium diboride possesses a comparatively high critical transition 
temperature compared with other conventional superconductors of about $T_c$=40~K. 
Presently only the high-$T_c$ cup\-rates have higher transition temperatures. 
For that reason the natural question arises whether superconductivity in 
MgB$_2$ is of the conventional electron-phonon driven type or if its 
superconducting state has more similarities with the one in the high-$T_c$ 
cuprates. In the high-$T_c$ cuprates the superconducting state is of an 
unconventional $d$-wave type, possessing gap nodes, and electronic pairing 
mechanisms, like for example exchange of antiferromagnetic spin fluctuations, 
are discussed seriously. An ongoing discussion still concerns the role and 
importance of phonons in these compounds. 

In contrast to optimally doped high-$T_c$ cuprates MgB$_2$ shows a strong 
isotope effect. If the boron-11 isotope is replaced by the lighter boron-10 
isotope, $T_c$ increases by about one Kelvin, indicating an important phonon 
contribution to the pairing interaction \cite{Budko}. Low temperature specific heat 
and penetration depth studies are consistent with an exponential decay, 
indicating the presence of a full gap without nodes \cite{Bouquet,Prozorov}.
In addition, 
there are no indications of sizeable magnetic interactions in MgB$_2$, again 
in contrast to the high-$T_c$ cuprates. All these observations seem to place 
MgB$_2$ among the conventional $s$-wave electron-phonon driven superconductors. 
Then the question arises, why it has such a high transition temperature as 
compared with other conventional superconductors. Here, band structure 
calculations turned out to be elusive, which we want to review in the following.

\begin{figure}
  \begin{center}
    \includegraphics[width=0.5\columnwidth,angle=270]{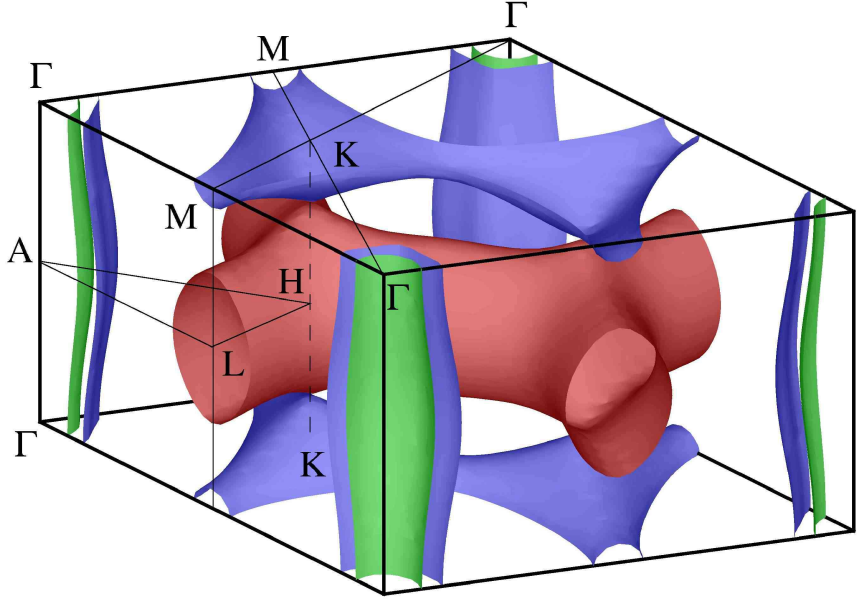}
    \vspace{.1cm}

    \caption{ The Fermi surface of MgB$_2$. (Adapted from Ref. \cite{Kortus1},
\copyright 2001 The American Physical Society).
    \label{Fermisurface}}
  \end{center}
\end{figure} 

The MgB$_2$ lattice structure consists of alternating layers of Boron and 
Magnesium atoms. The Boron atoms form a honeycomb lattice and the Magnesium 
atoms a triangular lattice halfway between the Boron layers. Calculations of 
the electronic band structure show four bands crossing the Fermi energy leading 
to four topologically disconnected Fermi surface sheets shown in 
Fig.~\ref{Fermisurface} \cite{Kortus1}. 
Two of these bands are derived from Boron p$_z$ orbitals. 
They form the so-called $\pi$ bands seen as the red (electronlike) and 
blue (holelike) tubular networks in Fig.~\ref{Fermisurface}. The other two bands derive from 
Boron p$_x$ and p$_y$ orbitals and form the 
so-called $\sigma$ bands, seen as the green and blue cylindrical Fermi surfaces 
centered around the $\Gamma$ point (both holelike). These possess mainly two 
dimensional character. Interestingly, all these bands are dominated by Boron p
orbitals and contributions from Magnesium orbitals are very small at the Fermi 
level. About 58 percent of the total density of states at the Fermi level is 
residing on the $\pi$ bands making both $\sigma$ and $\pi$ bands about equally 
important for the electronic properties of MgB$_2$.
            
Density-functional calculations of the phonon modes and the electron-phonon 
interaction strength can be found in Refs. \cite{Andersen}, \cite{Bohnen}, 
and \cite{Kortus2}. The highest 
phonon density of states is found in the energy range around 30 meV. However, 
these phonons only couple weakly to the electrons at the Fermi level and thus 
do not contribute very much to superconductivity. This can be nicely 
recognized in Fig.~1 of Ref. \cite{Andersen}, where the interaction 
strength of the phonons is shown 
as the area of the black circles in the figure. In fact, the phonons that 
couple most strongly to the electrons at the Fermi level are found in the 
energy range around 70 meV. These phonon modes evolve from the E$_{2g}$ mode 
at the $\Gamma$  point and correspond to a Boron-Boron bond-stretching 
vibration of the Boron sub lattice. A comparison with the phonon modes in 
the isostructural but nonsuperconducting compound AlB$_2$ in Ref. 
\cite{Bohnen} shows that these E$_{2g}$ phonon modes 
are strongly softened in MgB$_2$ consistent with their strong coupling. 
Correspondingly, the so-called Eliashberg function $\alpha^2 F(\omega)$, 
which weights the 
phonon density of states with the coupling strengths and appropriately 
describes the pairing interaction due to phonons, possesses a strong peak 
around 70 meV and significantly differs from the phonon density of states in 
contrast to conventional strong-coupling superconductors. The dimensionless 
electron-phonon coupling constant was found to lie between 
$\lambda \approx$~0.7-0.9 from 
these first-principles calculations. From this microscopic information we can 
obtain a qualitative understanding of why $T_c$ is so high in MgB$_2$ by 
looking at the BCS $T_c$ formula (which is of course a bad approximation in 
the present case, but can give us some qualitative tendencies):
\begin{equation}
k_B T_c = 1.13 \hbar \omega_c e^{-1/V N(0)}
\end{equation} 
Here, $\omega_c$ is a characteristic phonon frequency, $V$ the interaction 
strength and $N(0)$ the density of states at the Fermi level, with 
$\lambda \sim VN(0)$. First of all, the characteristic phonon frequency is 
comparatively high, because Boron is a light element and the E$_{2g}$ phonon 
modes in question only involve vibrations of the Boron sub lattice. Secondly, 
this high frequency phonon at the same time possesses a strong coupling to the 
electrons at the Fermi level. This means that in MgB$_2$ we have a favorable 
coincidence of two effects helping to raise $T_c$.
\vspace{2ex}

\noindent
{\bf 3. TWO GAP SUPERCONDUCTIVITY IN MAGNESIUM DIBORIDE}
\vspace{1ex}

It was noted already early on that there is a problem with the superconducting 
gap size in MgB$_2$. From the BCS gap ratio $\Delta_0=1.76 k_B T_c$ one should 
expect a gap of 6 meV or somewhat larger, if strong electron-phonon coupling 
effects are considered. However, experimental results obtained from different 
experimental techniques seemed to scatter between 2 meV and 8 meV with some 
clustering around 2.5 meV and 7 meV \cite{Buzea}. Initially, one might have 
attributed 
this to insufficient sample quality, however, by the end of 2001 high quality 
single crystals became available and this problem remained. Also, values of 
the coherence length $\xi$ extracted by different means turned out to differ 
considerably. For example, STM tunneling measurements of the vortex core size 
by Eskildsen et al. \cite{Eskildsen} were consistent with a value of 
$\xi \approx$50~nm, 
while the coherence length as determined from the upper critical field value 
on the same sample gave $\xi \approx$10~nm. Since the coherence length is 
related to the superconducting gap via $\xi=\hbar v_F/\pi \Delta_0$ these 
differences can be related to differences in the gap value as well. These 
problems suggested studying the possibility of anisotropic or multiple gap 
structures in MgB$_2$.

Again, important insight into this question was provided by Migdal-Eliashberg 
type calculations based on band structure calculations 
\cite{Kortus2,Golubov,ChoiB,ChoiN}. 
Migdal-Eliashberg theory is a generalization of BCS-theory and allows to take
into account the detailed properties of phonons and their coupling to the
electrons at the Fermi level in the pairing interaction.
Decomposing the electron-phonon coupling 
constant into contributions from the four bands Liu et al. \cite{Kortus2} were able to 
show that the pairing interaction differs considerably on the $\sigma$ and 
$\pi$ bands. The pairing interaction on the $\sigma$ bands turns out to be 
much stronger than the one on the $\pi$ bands and the interband pairing strength. 
The reason for this is that the E$_{2g}$ phonon mode being an in-plane vibration 
of the Boron atoms preferentially couples to the in-plane electrons on the more 
two dimensional $\sigma$ bands, while its coupling to the three dimensional 
$\pi$ bands is much weaker. Due to this anisotropy the resulting effective 
coupling constant was shown to increase to about $\lambda \approx 1$, being 
quantitatively more consistent with a $T_c$ of 40 K.

\begin{figure}[t]
  \begin{center}
    \includegraphics[width=0.4\columnwidth,angle=0]{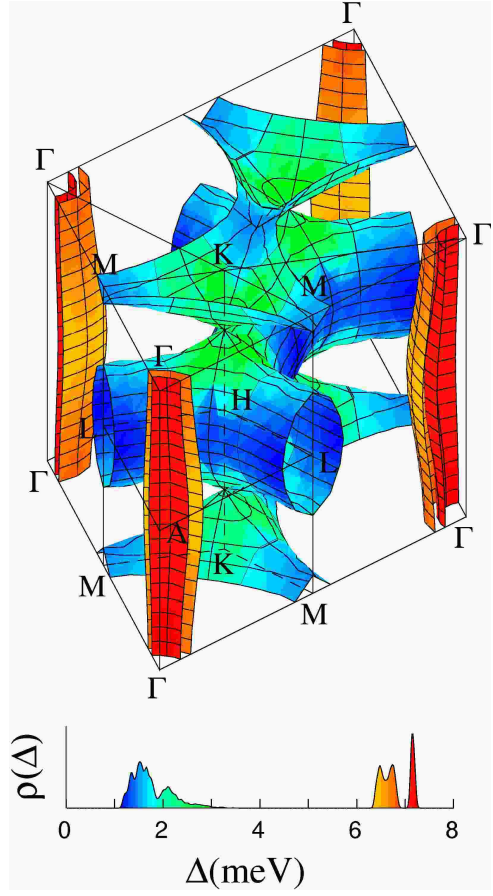}
    \vspace{.1cm}

    \caption{ The energy gap distribution on the Fermi surface of MgB$_2$ 
(color coded). (Adapted from Ref. \cite{ChoiN},
\copyright 2002 Nature Publishing Group).
    \label{gapdistr}}
  \end{center}
\end{figure} 

A fully momentum dependent solution of Migdal-Eliashberg equations was 
presented by H.~J.~Choi et al in Refs. \cite{ChoiB} and \cite{ChoiN}. 
In this calculation the fully 
momentum dependent band structure, phonon modes, and electron-phonon couplings 
obtained from first principles density functional calculations were used as a 
starting point to solve Migdal-Eliashberg equations in the full Brillouin zone. 
The Coulomb pseudopotential was taken as a constant $\mu^*=0.12$, which 
reproduces the experimental $T_c$. Neglecting impurity scattering, this 
procedure allowed to calculate the value of the superconducting gap at each 
point on the Fermi surface. The result is shown in Fig.~\ref{gapdistr}, 
where the size of 
the gap is shown color coded on top of the Fermi surface structure. This 
picture impressively demonstrates the presence of a large gap value around 
7 meV on the $\sigma$ bands and a small gap value around 2 meV on the $\pi$ 
bands.

Consistent with Ref. \cite{Kortus2} the authors were able to demonstrate the important 
role of both anisotropy and anharmonicity of the phonons on $T_c$. When 
anisotropy was neglected and only the isotropic Migdal-Eliashberg equations 
were solved, $T_c$ dropped from 39 K to 19 K. When anharmonicity on the phonon 
frequencies was neglected but anisotropy was kept, $T_c$ was seen to increase 
from 39 K to 55 K \cite{ChoiB}. This shows that phonon anharmonicity is harmful for 
superconductivity in MgB$_2$, but anisotropy is helpful, and for a 
quantitative understanding of superconductivity in MgB$_2$ both are needed. 
Choi et al. were also able to show that the isotope exponent observed 
experimentally could be quantitatively reproduced, if the anharmonicity of the 
phonons was taken into account. Also, the temperature dependence of the 
specific heat from these calculations was in good agreement with the 
experimental results.
All these results seem to show that we can obtain a complete understanding in 
terms of two gap superconductivity in MgB$_2$ based on the picture shown 
in Fig.~\ref{gapdistr}.

An important question in this picture concerns the role of impurities and in 
particular interband impurity scattering, however. Conventionally one would 
expect that impurity scattering tends to equalize all gap values on the Fermi 
surfaces thereby reducing $T_c$. This is one of the reasons why two gap 
superconductivity or anisotropic $s$-wave superconductivity is rarely observed. 
This does not seem to be the case in MgB$_2$. For example, samples with very 
different residual resistivities are observed to have essentially the same 
critical temperature $T_c$. This is a behavior that one would expect in a 
single gap $s$-wave superconductor due to Anderson's theorem. However, in a 
two gap superconductor a strong change of $T_c$ with impurity concentration 
should be expected \cite{Schopohl}. A very good answer to this question has 
been given by 
I.~I.~Mazin et al. in Ref. \cite{Mazin}: due to the particular electronic structure 
of MgB$_2$ the electronic wave functions on $\pi$ bands and $\sigma$ bands 
possess different parity symmetry. The $\pi$ bands deriving from the Boron 
p$_z$ orbitals are mainly antisymmetric with respect to the Boron plane, 
while the $\sigma$ bands deriving from both Boron p$_x$ and p$_y$ orbitals 
are mainly symmetric. This disparity of $\sigma$ and $\pi$ bands makes the 
impurity scattering matrix element between these two types of bands 
exceptionally small as compared with the impurity scattering matrix element 
within each of these bands. Using density functional supercell calculations 
for various impurities Mazin et al were able to show that the interband 
scattering rate appears to be one to two orders of magnitude smaller than the 
intraband scattering rates due to this band disparity. This means that 
impurity scattering will equalize the gaps within each of the two types of 
bands, but not very much between them. This makes the two gaps in MgB$_2$ 
exceptionally stable against impurity scattering. Only a very large amount 
of impurity scattering is expected to lead to a reduction of $T_c$ due to 
interband scattering and an accompanied merging of the two gaps.

While signatures of the two gaps have been observed in several different
experiments, so far direct experimental confirmation of the important role
of the E$_{2g}$ phonon for the pairing is still lacking. In principle,
such a strong coupling phonon could produce observable structures in the
tunneling density of states. However, there are difficulties in the case 
of MgB$_2$ as has been discussed by Dolgov et al \cite{Dolgov}: on one 
hand the
high frequency of the phonon mode leads to a strong reduction of the
structures produced in the tunneling spectrum, on the other hand
tunneling spectroscopy is mostly dominated by the $\pi$ band phonons,
which are not so important for the pairing interaction. This makes
observation of the E$_{2g}$ phonon and its coupling strength in 
tunneling spectroscopy difficult.

Concluding this section we can say that we seem to have a consistent and 
quantitative picture of two gap superconductivity in MgB$_2$, which mainly 
arises from density functional calculations and solutions of Migdal-Eliashberg 
equations. The two gaps arise due to a strongly anisotropic electron-phonon 
interaction of the (anharmonic) E$_{2g}$ phonon mode, which couples more 
strongly to the electrons on the $\sigma$ bands than to the ones on the 
$\pi$ bands. Due to the disparity of the two types of bands these two gaps 
appear to be very stable against impurity scattering, allowing $T_c$ to 
remain large even in samples with large residual resistivities.
\vspace{2ex}

\noindent
{\bf 4.  UPPER CRITICAL FIELD ANISOTROPY}
\vspace{1ex}

In this section we want to explore the consequences of the two gap picture 
presented in the previous section on the temperature dependence of the upper 
critical field $B_{c2}$, particularly its anisotropy. Measurements of the 
upper critical field in MgB$_2$ single crystals have shown that not only 
$B_{c2}$ is anisotropic, but also that this anisotropy is strongly 
temperature dependent \cite{Angst,Eltsev,Lyard,Welp}. 
The anisotropy ratio 
$\Gamma_{c2}$ is given by $\Gamma_{c2}=B_{c2}^{ab}/B_{c2}^{c}$, where 
$B_{c2}^{ab}$ is the upper critical field, when the field is applied along 
the Boron planes, and $B_{c2}^{c}$ the one for field along $c$-axis direction 
perpendicular to the Boron planes. At low temperatures $\Gamma_{c2}$ reaches 
values around 5. It decreases with increasing temperature and reaches values 
around 2 at $T_c$ (see the solid circles in Fig.~\ref{Fig2DS} below). 
Close to $T_c$ $B_{c2}^{c}$ varies linearly with 
temperature, while $B_{c2}^{ab}$ shows a pronounced upward curvature. 
This behavior is quite unusual, because in 
conventional single gap superconductors the anisotropy ratio is very much 
temperature independent, rarely changing by more than 10 to 20 percent. 
An upward curvature of the upper critical field has been observed in 
YNi$_2$B$_2$C and attributed to two band behavior, however, 
in this case the upward curvature appeared in all spatial directions, in 
contrast to MgB$_2$ \cite{Shulga}. 

The first theoretical work addressing this unusual behavior in MgB$_2$ was a 
study within an anisotropic gap model by Posazhennikova et al. \cite{Posazh}.
In this work a single anisotropic $s$-wave gap on an elliptical Fermi 
surface was 
considered. It was shown that the experimental data of the upper critical 
field in MgB$_2$ single crystals including both the strong temperature 
dependence of the anisotropy ratio as well as the upward curvature only 
appearing with field in $ab$-plane direction could be fitted within this 
model, 
if the gap was smaller in $c$-axis direction than in $ab$-plane direction. 
While this model can account for all experimental observations on the upper 
critical field in MgB$_2$, it possesses some drawbacks, however, regarding 
its consistency with other experimental results. One problem concerns the 
ratio of 
the maximum to the minimum gap value. Within this anisotropic $s$-wave model 
a gap ratio of about 1:10 was needed to explain the upper critical field data. 
This ratio seems too large compared with the about 1:3 ratio as observed in 
tunneling experiments, for example. In addition, measurements of the in-plane 
penetration depth clearly show an exponential decrease with the small 
gap \cite{JinPRB,DPM}. This suggests that the small gap is present within the 
Boron plane direction consistent with the picture in Fig.~\ref{gapdistr}, 
but inconsistent with this anisotropic $s$-wave model.

Another problem concerns the anisotropy of the lower critical field $B_{c1}$. 
From
anisotropic (single gap) Ginzburg-Landau theory one should expect that the 
upper critical field anisotropy is related to the lower critical field 
anisotropy via $B_{c2}^{ab}/B_{c2}^{c}=B_{c1}^{c}/B_{c1}^{ab}$. Recent 
experimental studies have 
established, however, that this relation is violated in MgB$_2$ 
\cite{Lyard2,Cubitt,MXu}, 
the upper critical field anisotropy decreasing with increasing temperature, 
while the lower critical anisotropy is found to increase. 
This is in agreement 
with expectations based on two gap models \cite{Miranovic}, but again 
inconsistent with an anisotropic $s$-wave model.

\begin{figure}
  \begin{center}
    \includegraphics[width=0.8\columnwidth,angle=0]{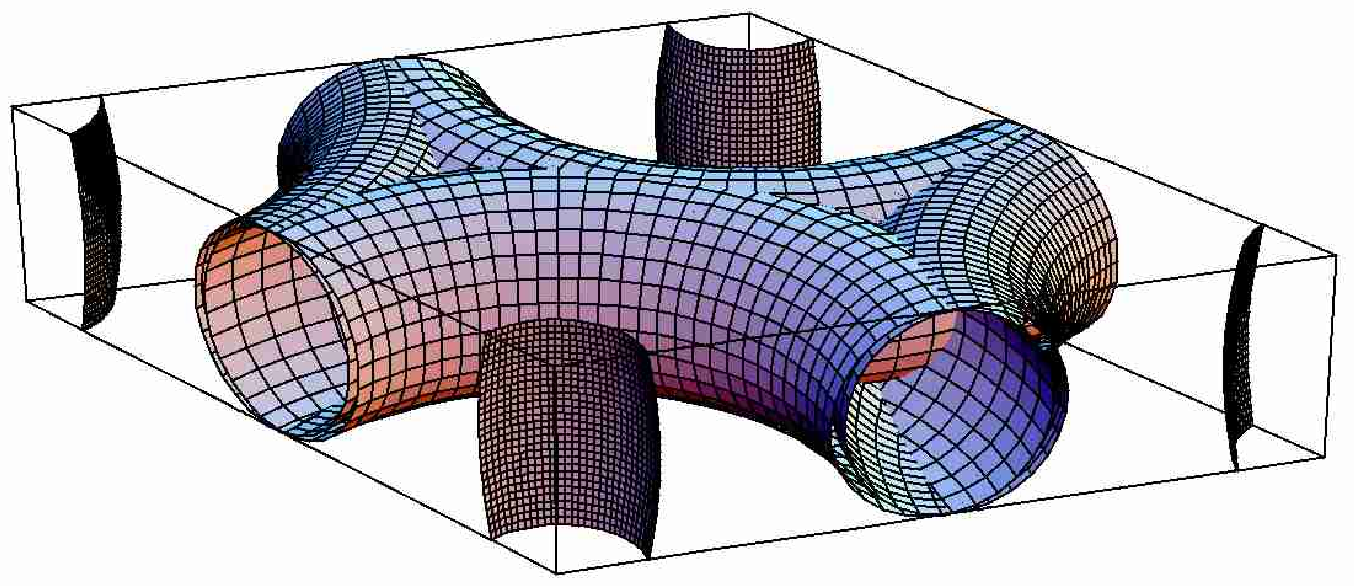}
    \includegraphics[width=0.95\columnwidth,angle=0]{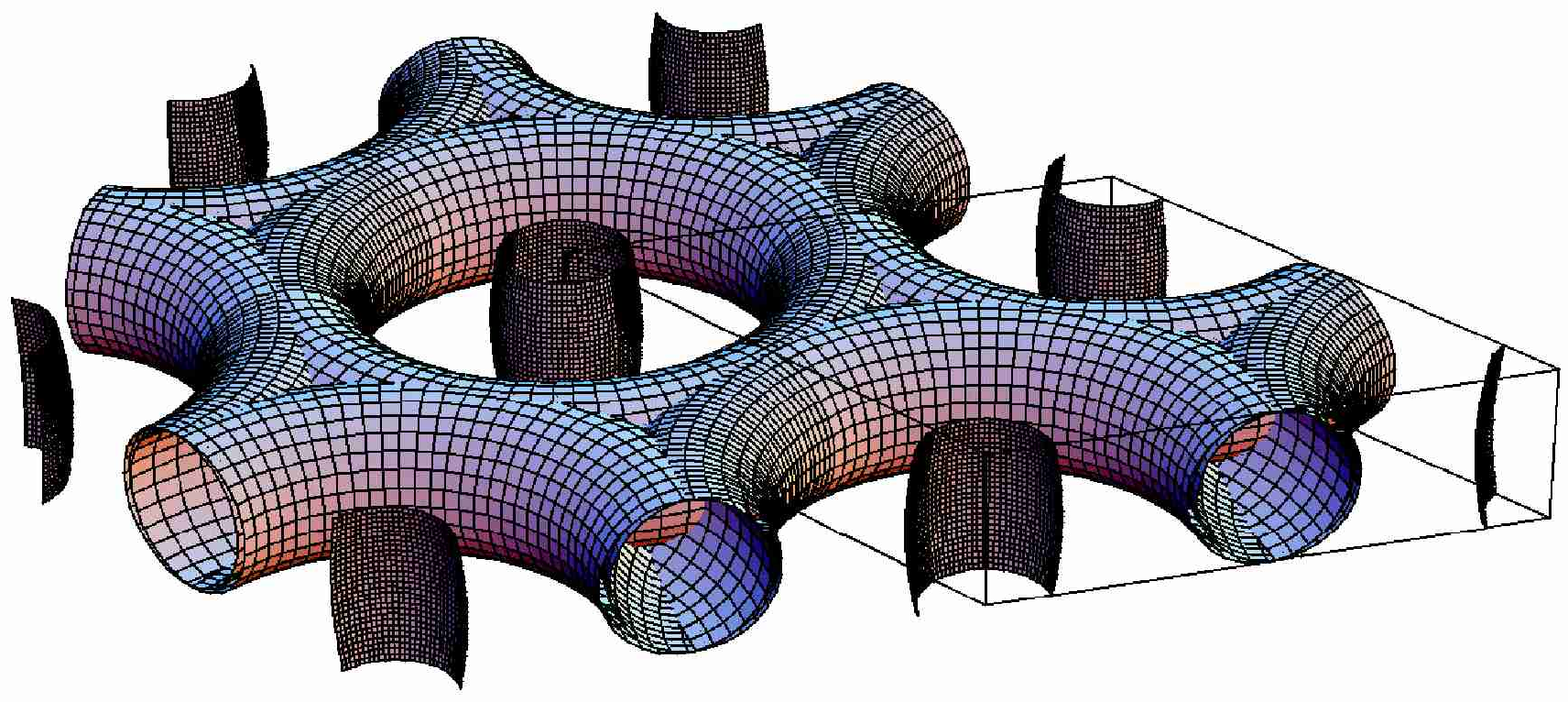}
    \includegraphics[width=0.95\columnwidth,angle=0]{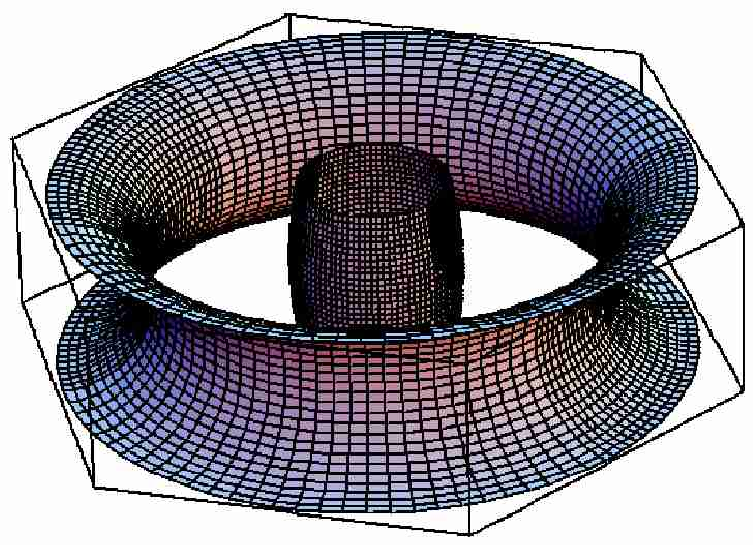}
    \vspace{.1cm}

    \caption{Simplification of the MgB$_2$ Fermi surface from 
    Fig.~\ref{Fermisurface}. 
Rearranging the Fermi surface sheets around the $\Gamma$ point shows
that the $\sigma$~band can be modeled as a distorted cylinder and the
$\pi$~band as a half-torus.
    \label{modelfermisurface}}
  \end{center}
\end{figure} 

A first calculation within an anisotropic two band model was presented by 
Miranovic et al. \cite{Miranovic} based on clean limit Eilenberger theory. 
The Fermi 
surface sheets of the two bands were taken as ellipsoids with very different 
anisotropies, while the two gaps were taken to be isotropic. Within this 
calculation it was also possible to qualitatively reproduce the experimentally 
observed upper critical field anisotropy as in the anisotropic $s$-wave model 
by Posazhennikova et al. In addition to that, however, this two band 
calculation did not suffer from the drawbacks mentioned above: more realistic 
values for the ratio of the two gaps could be taken, the small gap was present 
within the Boron plane direction, and the lower critical field anisotropy was 
found to increase with temperature.

In order to answer the question whether the picture in Fig.~\ref{gapdistr} 
is consistent 
with the upper critical field anisotropy the author and N.~Schopohl presented 
a calculation within two band Eilenberger theory taking into account the 
detailed band structure Fermi surface topology based on 
Fig.~\ref{Fermisurface} \cite{DahmSchopohl}. For that purpose a simple
but realistic model of the Fermi surface was used
as shown in Fig.~\ref{modelfermisurface}. The upper panel of
Fig.~\ref{modelfermisurface} shows one half
of the Brillouin zone from Fig.~\ref{Fermisurface}. Using the periodicity
of the lattice identical copies of the Brillouin zone may be attached to
each other (center panel). Choosing a new Brillouin zone around the
$\Gamma$ point leads to the structure shown in the lower panel.
This rearrangement shows that the $\sigma$~band Fermi surface can be 
modeled by a distorted cylinder and the $\pi$~band Fermi surface by
a torus cut open half. Here, the two $\sigma$~bands and the two
$\pi$~bands are assumed to be degenerate, for simplicity. This suggests 
the following parameterization
of the Fermi surfaces for the $\sigma$~band Fermi momentum \cite{GDS}:
\begin{equation}
\vec{k}_{F,\sigma} \left( \varphi, k_z \right) = \left( \begin{array}{c} 
\left( k_{F,\sigma} + \frac{\epsilon_c}{c} \cos \left( c k_z \right) \right)
\cos \varphi \\
\left( k_{F,\sigma} + \frac{\epsilon_c}{c} \cos \left( c k_z \right) \right)
\sin \varphi \\ k_z \end{array} \right)
\end{equation}
and the $\pi$~band Fermi momentum:
\begin{equation}
\vec{k}_{F,\pi} \left( \varphi, \vartheta \right) = \left( \begin{array}{c} 
k_{F,\pi} \left( \frac{1}{\kappa} + \cos \vartheta \right) \cos \varphi \\
k_{F,\pi} \left( \frac{1}{\kappa} + \cos \vartheta \right) \sin \varphi \\
k_{F,\pi} \sin \vartheta
\end{array} \right) 
\end{equation}
Here, $\varphi$ is the in-plane angle, $k_z$ the $c$-axis momentum, and
$\vartheta$ the azimuthal angle of the half-torus running from
$\pi/2$ to $3\pi/2$. The corresponding Fermi velocities are given by
\begin{equation}
\vec{v}_{F,\sigma} \left( \varphi, k_z \right) = v_{F,\sigma} 
\left( \begin{array}{c} 
\cos \varphi \\ \sin \varphi \\ \epsilon_c \sin c k_z \end{array} \right)
\quad {\mathrm{and}} \quad
\vec{v}_{F,\pi} \left( \varphi,  \vartheta \right) = v_{F,\pi} 
\left( \begin{array}{c} 
\cos \varphi \cos \vartheta \\ \sin \varphi \cos \vartheta 
\\ \sin \vartheta \end{array} \right)
\label{Fermivel}
\end{equation}
Here, the dimensionless $c$-axis dispersion parameter $\epsilon_c$
has been assumed small. For MgB$_2$ the following parameter values
have been extracted from band structure calculations \cite{DahmSchopohl}:
$v_{F,\sigma}=4.4 \times 10^5$~m/s, $v_{F,\pi}=8.2 \times 10^5$~m/s,
$\epsilon_c=0.23$, and $\kappa=0.25$.

Based on this model for the MgB$_2$ Fermi surface one can now
calculate the upper critical field using
the linearized two band Eilenberger equations. For $\omega_n>0$
they read
\begin{equation}
\left\{ \omega_n +
\vec{v}_{F,\alpha} \left[
\frac{\hbar}{2} \vec{\nabla} - i \frac{e}{c} \vec{A} \left( \vec{r} \right)
\right] \right\}  f_{\alpha} (\vec{r}, \hat{k}; \omega_n) =
- \Delta_\alpha (\vec{r})
\label{Eilenbergereq} 
\end{equation}
along with the gap equation
\begin{equation}
\Delta_\alpha (\vec{r}) = - \pi T
\sum_{\alpha'} \sum_{|\omega_n^\prime| < \omega_c} \lambda^{\alpha \alpha'} 
\left\langle
f_{\alpha'} (\vec{r}, \hat{k}^\prime; \omega_n^\prime)
\right\rangle_{\alpha'}
\label{gapeqnbc2}
\end{equation}
Here, $f_{\alpha}$ is the anomalous Eilenberger propagator and
$\Delta_\alpha (\vec{r})$ the (spatially dependent) gap function
for the two bands $\alpha \in \{ \pi,\sigma \}$. The pairing interaction
$\lambda^{\alpha \alpha'}$ becomes a two-by-two matrix in the band
indices. The effective matrix elements have been calculated from
band structure calculations in Ref.~\cite{Kortus2} and found to be
\begin{equation}
\left( \begin{array}{cc} 
\lambda^{\sigma \sigma} & \lambda^{\sigma \pi} \\
\lambda^{\pi \sigma} & \lambda^{\pi \pi}
\end{array} \right) =
\left( \begin{array}{cc} 
0.959 & 0.222 \\ 0.163 & 0.278
\end{array} \right).
\label{lambdamat}
\end{equation} 
Together Equations (\ref{Eilenbergereq}) and (\ref{gapeqnbc2})
establish an eigenvalue problem for the gap function $\Delta_\alpha (\vec{r})$.
For a given temperature $T$ one has too look for the solution 
$\Delta_\alpha (\vec{r})$ that solves both equations for the highest
value of the magnetic field $\vec{B}=\vec{\nabla} \times \vec{A}$.

Eilenberger theory
is a generalization of BCS theory to inhomogeneous superconducting
states and contains Ginzburg-Landau theory as a limiting case
for $T \rightarrow T_c$ \cite{Eilenberger,LarkinOv}. 
It holds in the limit $k_F \xi \ll 1$,
where $\xi$ is the coherence length and $k_F$ the Fermi
momentum. Since Ginzburg-Landau theory is limited to the
vicinity of $T_c$, Eilenberger theory is the method of choice,
if one wants to calculate properties of type-II superconductors
in the vortex state at lower temperatures from microscopic grounds.
Near the upper critical field the gap function becomes small and
therefore one can use the linearized Eilenberger equations to
determine $B_{c2}$. 

Equations (\ref{Eilenbergereq}) and (\ref{gapeqnbc2}) are usually
solved using a Landau level expansion of $\Delta_\alpha (\vec{r})$ 
above the Abrikosov ground state of the vortex lattice \cite{Rieck}.
For strongly anisotropic systems this procedure requires a high
number of excited states, however. In addition, the calculation can
only be done numerically.
In Ref.~\cite{DahmSchopohl} Equations (\ref{Eilenbergereq}) and (\ref{gapeqnbc2})
were solved using a variational method that had been introduced earlier
by the author in Ref.~\cite{Posazh}. This method does not require
a high number of excited states and even allows to obtain analytical
results in some limiting cases. The variational ansatz for 
$\Delta_\alpha (\vec{r})$ 
corresponds to a distorted Abrikosov ground state of the form
\begin{equation}
\Delta_\alpha (\vec{r})= \Delta_\alpha \psi_\Lambda \left( e^{-\tau} x,
e^\tau y \right)
\label{varground}
\end{equation}
where $\tau$ is used as a variational parameter and determined such
as to maximize the upper critical field $B_{c2}$. Here, $\psi_\Lambda$
is the usual (undistorted) Abrikosov ground state. More
details about this variational method can be found in the
appendix.

Using this method the temperature dependence of the anisotropy ratio
was calculated from microscopic grounds based on band structure
calculations in Ref.~\cite{DahmSchopohl} as pointed out above. 
The result is shown in Fig.~\ref{Fig2DS}. Here, the parameter
$\eta$ is a dimensionless parameter describing the interband
pairing strength and is given by 
\begin{equation}
\eta=\frac{\lambda^{\pi \pi} - \lambda_-}{\lambda_+ - \lambda_-}
\label{defeta}
\end{equation}
where $\lambda_+$ and $\lambda_-$ are the larger and smaller
eigenvalues of the matrix $\lambda^{\alpha \alpha'}$ in
Eq.~(\ref{gapeqnbc2}), respectively. The dashed line shows the result
for the parameters obtained from band structure calculations,
the solid circles are experimental results on MgB$_2$ single
crystals from Lyard et al.~\cite{Lyard}. For comparison,
also the result with no interband pairing $\eta=0$ is
shown as the solid line. In this case the upper critical
field anisotropy is determined by the $\sigma$ band only.

\begin{figure}
  \begin{center}
    \includegraphics[width=0.42\columnwidth,angle=270]{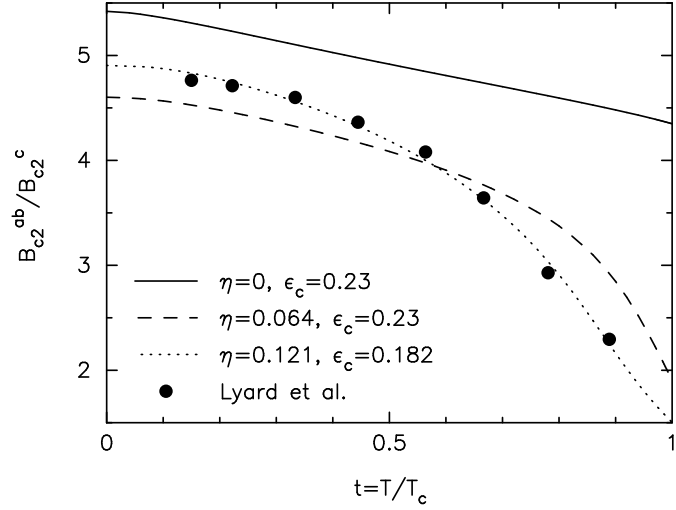}
    \vspace{.1cm}

    \caption{Temperature dependence of the anisotropy ratio
     $\Gamma_{c2}=B_{c2}^{ab}/B_{c2}^c$ for the band structure based
      two-band model of Ref.~\cite{DahmSchopohl}.
     Results are shown for different interband pairing
     strengths $\eta$ and $c$-axis dispersions $\epsilon_c$. Solid circles 
     are experimental results
     from Lyard et al.~\cite{Lyard} (adapted from Ref.~\cite{DahmSchopohl},
\copyright 2003 The American Physical Society).
     \label{Fig2DS}}
  \end{center}
\end{figure} 

In Ref.~\cite{DahmSchopohl} it was found that the two most important
parameters determining the temperature dependence of the upper
critical field anisotropy are $\eta$ and the $c$-axis dispersion
$\epsilon_c$ of the cylindrical $\sigma$ band. All other parameters
have only minor influence on the shape of the curves in Fig.~\ref{Fig2DS}.
If these two parameters are allowed to vary from their band structure
values, one can obtain an excellent fit of the experimental data
shown as the dotted line in Fig.~\ref{Fig2DS}. In this (two parameter)
fit the interband pairing strength $\eta$ turns out to be somewhat 
larger and the $c$-axis dispersion $\epsilon_c$ a little bit smaller
than expected from band structure calculations. Similar observations,
that larger interband pairing strengths than found by 
band structure calculations are needed to fit experimental
data, have been made also in Refs.~\cite{JinPRL} and \cite{Rydh}.
Possible explanations could be either that band structure calculations
underestimate the interband pairing strength or that some small
amount of interband impurity scattering or effects of strong 
electron-phonon coupling, which were neglected in the above 
calculations of the upper critical field, mimic a somewhat stronger
interband pairing strength. 

The result in Fig.~\ref{Fig2DS} shows that indeed a strong temperature
dependence of the upper critical field ratio consistent with experimental
results has to be expected
in the clean limit for the parameters found in band structure
calculations for MgB$_2$. As the solid line shows, the influence
of both gaps is crucial here. The strong temperature dependence
can be understood as follows: at low temperatures and high magnetic
fields the $\sigma$ band with the large gap is dominating leading
to a strong anisotropy. When temperature is increased and thus
the magnetic field reduced, the influence of the $\pi$ band
with the small gap becomes more and more important. Since
the $\pi$ band is more isotropic a reduction of the upper critical
field anisotropy results.

\begin{figure}
  \begin{center}
    \includegraphics[width=0.42\columnwidth,angle=270]{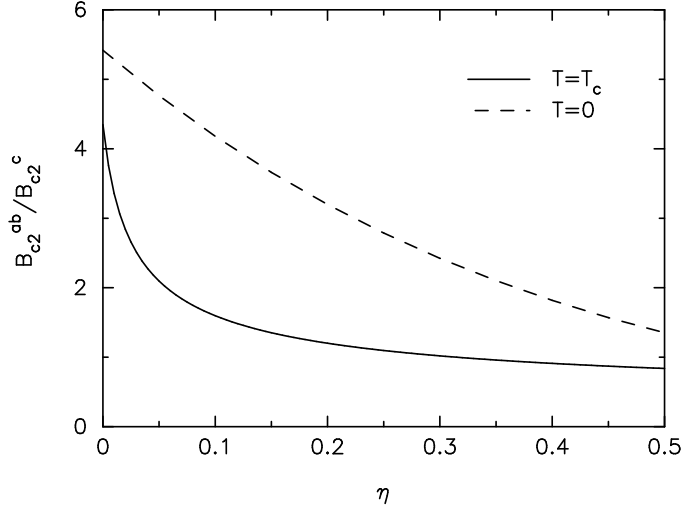}
    \vspace{.1cm}

    \caption{Anisotropy ratio
     $\Gamma_{c2}=B_{c2}^{ab}/B_{c2}^c$ as a function of interband pairing
     strength $\eta$ for the band structure based
      two-band model of Ref.~\cite{DahmSchopohl} at $T=0$ (dashed line)
      and $T=T_c$ (solid line)
      (adapted from Ref.~\cite{DahmSchopohl},
\copyright 2003 The American Physical Society).
     \label{Fig3DS}}
  \end{center}
\end{figure}

It is instructive to look at the change of $\Gamma_{c2}$ as a 
function of the interband pairing strength $\eta$ at $T=0$ and
$T=T_c$. This is shown in Fig.~\ref{Fig3DS}. When $\eta=0$
there is no strong temperature dependence of $\Gamma_{c2}$.
When $\eta$ is increased, $\Gamma_{c2}$ decreases much
more rapidly at $T=T_c$ than at $T=0$, because the influence
of the $\pi$ band is more important at higher temperature. 
As a result the temperature dependence becomes stronger.
However, when $\eta$ is increased towards 0.5 (maximum
coupling of the two bands), eventually $\Gamma_{c2}$
becomes small also at low temperature and the temperature
dependence becomes weak again. It appears that either
for very weak or for very strong interband pairing interaction 
the system effectively behaves like a single gap superconductor
with a weak temperature dependence of the upper critical
field anisotropy. It is only in the intermediate regime
around $\eta \sim 0.05-0.3$ where a strong temperature
dependence of $\Gamma_{c2}$ can be observed. This is
just the regime in which MgB$_2$ appears to be according
to band structure calculations.

In Ref.~\cite{DGS} further results of this model can be found:
the temperature dependence of the upper critical field for
field applied along the Boron plane direction shows an
upward curvature, while for field applied in $c$-axis direction
it does not, in good agreement with the experiments. Also, the 
magnetic field dependence of
the zero energy density of states follows the experimental
temperature dependence of the electronic specific heat 
coefficient.

The theories about the upper critical field in MgB$_2$ discussed so
far are clean limit theories, e.g. impurity scattering was assumed
to be small. However, even in the best MgB$_2$ single crystals
available so far it is believed that in the $\pi$ band the scattering
rate is larger than the gap, while in the $\sigma$ band the scattering
rate might be smaller than the gap. This means that impurities are
expected to affect the temperature dependence of the upper critical
field and one might ask how this could affect the clean limit results
discussed above. So far no calculations for general impurity scattering
rates have been presented. However, calculations can be considerably
simplified in the dirty limit. In Refs.~\cite{Gurevich,Koshelev}
such dirty limit theories of the upper critical field for MgB$_2$ were
presented. The main simplification here is that the intraband scattering
rates $\Gamma_\pi$ and $\Gamma_\sigma$ are assumed to be larger than
the gaps, while the interband scattering rate is assumed to be
negligible due to the parity argument by 
Mazin et al. \cite{Mazin} discussed in the previous section.
Interestingly, the strong temperature dependence of the upper critical
field anisotropy was shown to exist also in this dirty limit \cite{Koshelev}.
The physical reason for this behavior is the same as in the clean limit:
an interband pairing interaction of intermediate strength and a
strong anisotropy of the Fermi velocities in the $\sigma$ band which
results in a strong anisotropy of the diffusivities in this band.
Therefore, the inclusion of impurities qualitatively does not change 
this behavior.

In Ref.~\cite{Koshelev} also the angular dependence of the upper critical
field was studied. It was shown that this angular dependence
shows strong deviations from the results expected from anisotropic 
Ginzburg-Landau theory. The reason for this is that the $c$-axis 
coherence lengths in the two bands strongly differ. As a result the
validity of Ginzburg-Landau theory is reduced to a very narrow
region near the critical temperature $T_c$ \cite{Koshelev2}.

To summarize this section we have seen that the unusually strong
temperature dependence of the upper critical field anisotropy
can be nicely understood in terms of the two gap model, if
the particular Fermi surface topology of MgB$_2$ is
taken into account. The essential ingredients for this effect
are two Fermi surfaces with very different anisotropies and
an interband pairing interaction with an intermediate strength.
Too weak or too strong interband pairing would result in effectively
single gap behavior. According to band structure calculations MgB$_2$ 
fulfils these requirements possessing a strongly anisotropic
cylindrical $\sigma$ band and a more isotropic toroidal $\pi$
band. Parameters from band structure calculations can even 
reproduce the upper critical field anisotropy quantitatively.
This shows that the behavior of the upper critical field is
consistent with the band structure picture presented in the
previous sections.

\vspace{2ex}

\noindent
{\bf 5.  MICROWAVE CONDUCTIVITY}
\vspace{1ex}

In this section we want to discuss the consequences of two gap 
superconductivity on the microwave conductivity in MgB$_2$, particularly its 
temperature dependence at fixed frequency. In Ref. \cite{JinPRL} the observation of an 
anomalous microwave conductivity peak in MgB$_2$ thin films has been reported 
and its interpretation in terms of two gap behavior shall be discussed here. 
We will first start with a review of conductivity peaks in conventional 
superconductors as well as in high-$T_c$ cuprates and then compare them with 
the results in MgB$_2$.

\begin{figure}
  \begin{center}
    \includegraphics[width=0.42\columnwidth,angle=270]{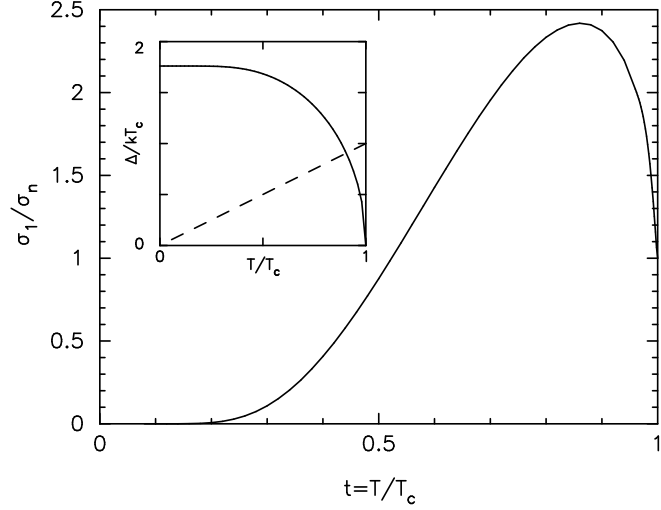}
    \vspace{.1cm}

    \caption{ Temperature dependence of the microwave conductivity at a 
frequency of $\omega=0.02 k_B T_c$ for a conventional $s$-wave 
superconductor in the dirty limit. 
Inset: temperature dependence of the gap (solid line) along with the 
line $\Delta=k_B T$ (dashed line). The crossing point roughly determines the 
position of the peak in the conductivity.
    \label{convcohpeak}}
  \end{center}
\end{figure} 

In conventional dirty limit superconductors the microwave conductivity at 
frequencies sufficiently below the gap frequency shows a coherence peak as a 
function of frequency, which is related to the so-called Hebel-Slichter peak 
in the temperature dependence of the NMR spin relaxation rate. As shown in 
Fig.~\ref{convcohpeak} the conductivity initially increases when entering 
the superconducting 
state going through a peak value near 0.9 $T_c$ and finally being suppressed 
exponentially at lower temperatures. The peak naturally comes out of BCS 
theory and at the time was regarded as one of the key triumphs of BCS theory, 
because earlier theories based on two fluid models were not able to account 
for this effect. When the superconducting state is entered and the gap opens, 
a square-root singularity appears in the density of states at the gap edge. 
This singularity leads to increased contributions to the microwave conductivity. 
When the gap $\Delta$ increases upon lowering the temperature, it eventually 
becomes larger than $k_B T$. At this point quasiparticles are condensed out into 
the superfluid and the microwave conductivity is suppressed exponentially at 
lower temperatures. As a rule of thumb we can estimate the position of the 
coherence peak roughly at the point where $\Delta(T)$ becomes equal to $k_B T$, 
which is shown as the crossing point between the solid and the dashed line in 
the inset of Fig.~\ref{convcohpeak}. In contrast to the NMR Hebel-Slichter 
peak, the 
coherence peak in the microwave conductivity depends on the scattering rate, 
however. When the system becomes cleaner, the microwave coherence peak is 
gradually reduced, as has been discussed by Marsiglio \cite{Marsiglio}. The ultimate 
reason for this dependence on the scattering rate is lying in the fact that 
the conductivity is a nonlocal quantity in contrast to the NMR relaxation 
rate which is a local probe.

\begin{figure}
  \begin{center}
    \includegraphics[width=0.5\columnwidth,angle=0]{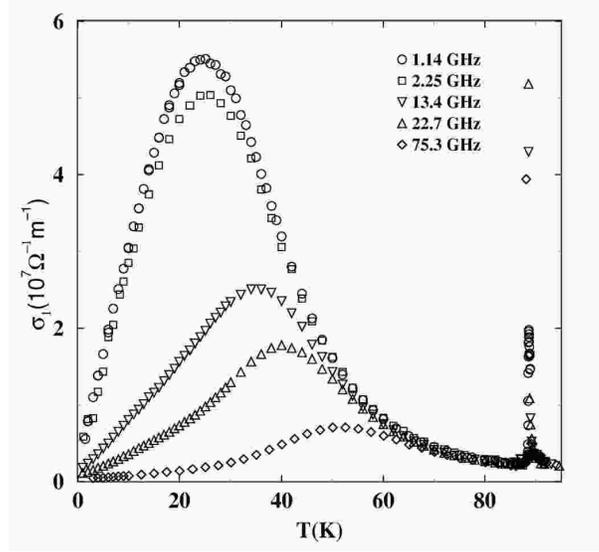}
    \vspace{.1cm}

    \caption{ Temperature dependence of the microwave conductivity in YBCO 
showing a large conductivity peak (adapted from Ref. \cite{Hosseini},
\copyright 1999 The American Physical Society).
    \label{condpeakhts}}
  \end{center}
\end{figure} 
                       
\begin{figure}
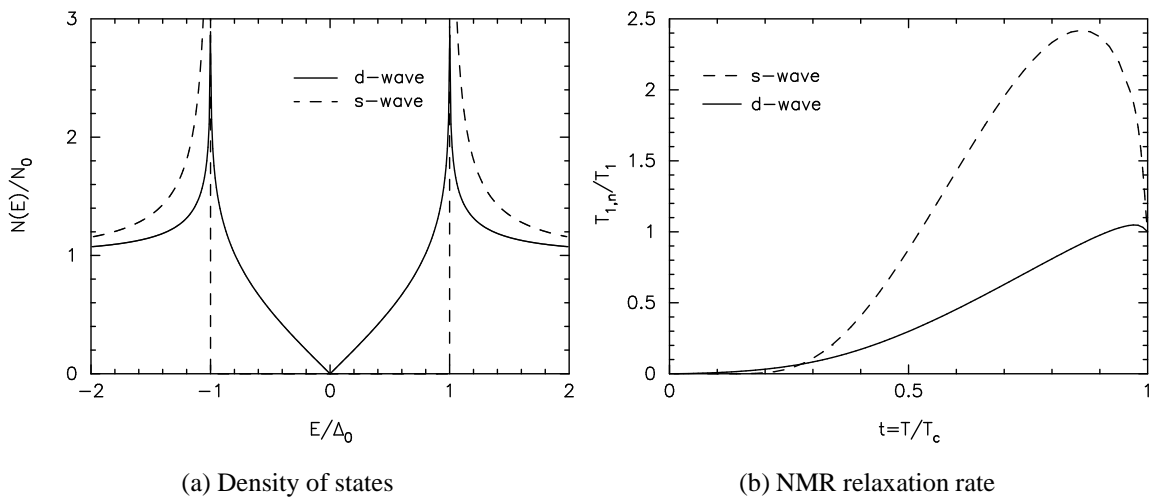

  \begin{center}
\subfigure[Density of states]{
    \includegraphics[width=0.36\columnwidth,angle=270]{dosds.ps}}
\subfigure[NMR relaxation rate]{
    \includegraphics[width=0.36\columnwidth,angle=270]{hebsli2.ps}}
    \vspace{.1cm}

    \caption{ (a) Density of states in a $d$-wave superconductor 
(solid line) 
and an $s$-wave superconductor (dashed line). (b) Corresponding 
NMR relaxation rates for $d$-wave (solid line) and $s$-wave (dashed line). 
In the $d$-wave case no large Hebel-Slichter peak is visible anymore.
    \label{dvss}}
  \end{center}
\end{figure}

In the high-$T_c$ cuprates the situation is completely different. In 
measurements of the NMR relaxation rate no Hebel-Slichter peak is observed. 
However, the microwave conductivity shows a huge peak at temperatures between 
0.3~$T_c$ and 0.6~$T_c$ depending on frequency as shown in 
Fig.~\ref{condpeakhts}. The absence 
of the Hebel-Slichter peak can be easily understood as a consequence of the 
$d$-wave nature of the superconducting state. In a $d$-wave superconductor 
the singularity at the gap edge of the density of states is not a square-root 
singularity anymore, but a logarithmic singularity, having a much weaker 
influence on the NMR relaxation rate, as shown in Fig.~\ref{dvss}. In addition, 
strong-coupling effects tend to wash out the singularities in the density of 
states, which leads to an additional suppression of the coherence peak as 
discussed in Ref. \cite{Marsiglio}.

Since there is no coherence peak in the NMR relaxation rate, why is there a 
peak in the microwave conductivity? It has been suggested that this peak has 
a different physical origin and is related to a rapid suppression of the 
inelastic scattering rate in the superconducting state \cite{Bonn}. 
Such a drop of 
the inelastic scattering rate naturally appears when inelastic scattering in 
the normal state is dominated by electron-electron scattering, for example by 
spin fluctuations, because the superconducting gap is suppressing 
electron-electron scattering below the gap energy in the superconducting state. 
Detailed theoretical calculations by Hirschfeld et al \cite{Hirschfeld} 
including $d$-wave 
superconductivity, strong impurity scattering, and inelastic spin fluctuation 
scattering were shown to be consistent with the experimental results.

\begin{figure}
  \begin{center}
    \includegraphics[width=0.7\columnwidth,angle=0]{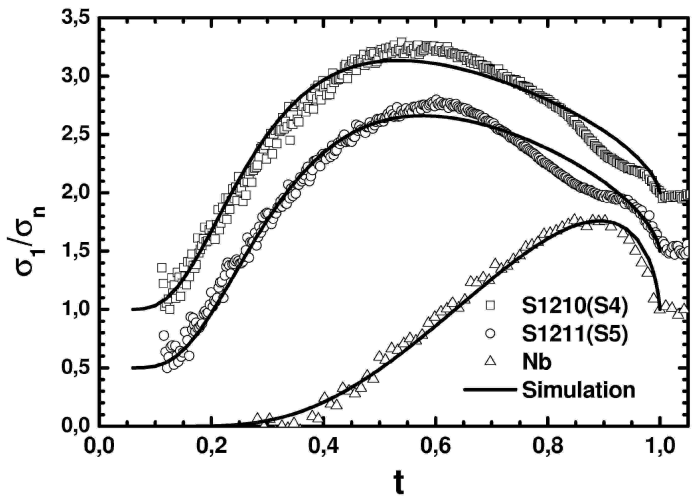}
    \vspace{.1cm}

    \caption{ Temperature dependence of the microwave conductivity in 
MgB$_2$ thin 
films (circles and squares). For comparison, the microwave conductivity of 
a Niobium thin film in the dirty limit is also shown (triangles). 
(Adapted from Ref. \cite{JinPRL},
\copyright 2003 The American Physical Society)
    \label{condpeakmgb2}}
  \end{center}
\end{figure} 

In MgB$_2$ microwave conductivity measurements have shown a conductivity peak 
appearing at about 0.5~$T_c$ as shown in Fig.~\ref{condpeakmgb2}. This peak 
position seems to 
be somewhat intermediate between conventional superconductors and high-$T_c$ 
cuprates and the natural question arises, whether this is a coherence peak or 
a peak due to lifetime effects as in the cuprates. In order to address the 
question of lifetime effects a quick look at the temperature dependence of 
the resistivity in MgB$_2$ is useful. It has been shown that the temperature 
dependence of the resistivity can be fitted well by Bloch-Gr\"uneisen formula 
with a temperature independent residual resistivity associated with impurity 
scattering and a temperature dependent phonon scattering part 
\cite{resistivity}. This seems 
to show that no large electron-electron scattering is present in the normal 
state of MgB$_2$. Moreover, at $T_c$ when superconductivity sets in, the 
resistivity is already in the saturation regime where phonon scattering is 
frozen out (due to the high frequency of the E$_{2g}$ phonon mode). Therefore 
the resistivity at $T_c$ seems to be dominated by elastic impurity scattering 
and thus no rapid drop of inelastic scattering can be expected in the 
superconducting state. Note that this is in contrast to the high-$T_c$ 
cuprates, where resistivity varies linearly with temperature down to $T_c$ 
and is not saturated. These arguments seem to exclude the interpretation of 
the conductivity peak in MgB$_2$ in terms of a lifetime effect.

In the following we therefore want to study the microwave conductivity of a 
dirty two gap superconductor. Dirty in the sense that the intraband impurity 
scattering rates in the two bands are assumed to be larger than the respective 
gaps, but interband scattering is neglected as suggested by the argument by 
Mazin et al. \cite{Mazin}.

\begin{figure}
  \begin{center}
\subfigure[]{
    \includegraphics[width=0.3\columnwidth,angle=0]{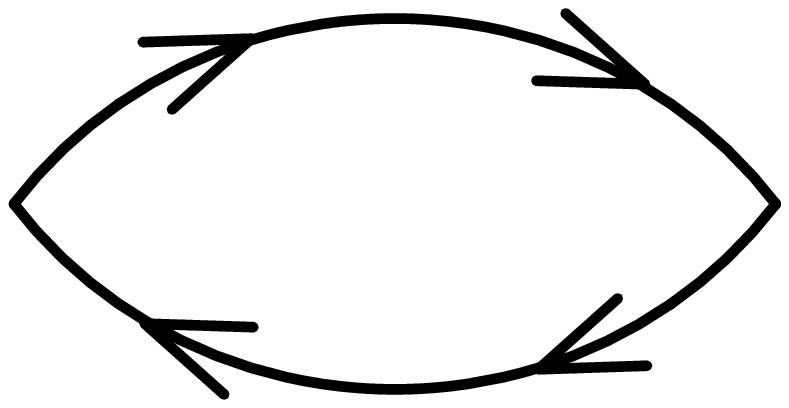}}
\begin{picture}(0,0)
\put(-125,10){\makebox(0,0){\Large\bf $\pi$}}
\put(-15,10){\makebox(0,0){\Large\bf $\pi$}}
\put(-125,57){\makebox(0,0){\Large\bf $\pi$}}
\put(-15,57){\makebox(0,0){\Large\bf $\pi$}}
\end{picture}
\subfigure[]{
    \includegraphics[width=0.3\columnwidth,angle=0]{bubble.eps}}
\begin{picture}(0,0)
\put(-125,10){\makebox(0,0){\Large\bf $\sigma$}}
\put(-15,10){\makebox(0,0){\Large\bf $\sigma$}}
\put(-125,57){\makebox(0,0){\Large\bf $\sigma$}}
\put(-15,57){\makebox(0,0){\Large\bf $\sigma$}}
\end{picture}
\subfigure[]{
    \includegraphics[width=0.3\columnwidth,angle=0]{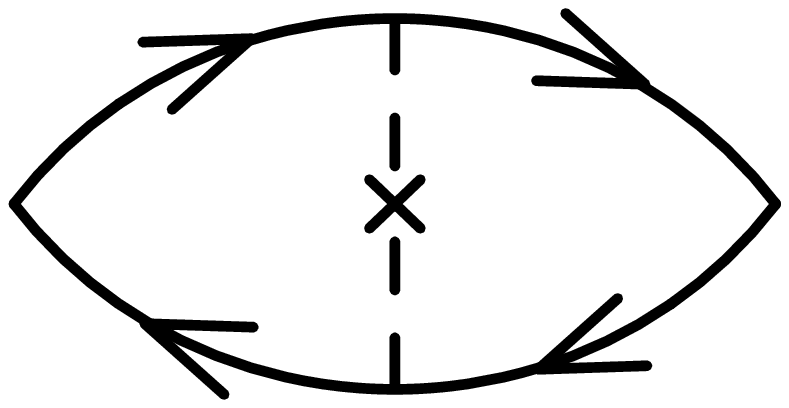}}
\begin{picture}(0,0)
\put(-125,10){\makebox(0,0){\Large\bf $\pi$}}
\put(-15,10){\makebox(0,0){\Large\bf $\sigma$}}
\put(-125,57){\makebox(0,0){\Large\bf $\pi$}}
\put(-15,57){\makebox(0,0){\Large\bf $\sigma$}}
\end{picture}
    \vspace{.1cm}

    \caption{ Diagrammatic contributions to the conductivity. 
(a) The main contribution to the $\pi$ band. (b) The main contribution to the $\sigma$ band.
(c) The lowest order interband scattering contribution to the conductivity which 
is expected to be small in MgB$_2$.
    \label{conddiagram}}
  \end{center}
\end{figure}

In a two band system like MgB$_2$ we can approximate the total conductivity 
as the sum of the two partial conductivities of the two bands (parallel 
resistor formula). This approximation neglects interband scattering events 
like the one shown diagrammatically in Fig.~\ref{conddiagram}(c). However, 
these events are expected to give only 
small corrections, because interband impurity scattering is expected to be 
much weaker than intraband impurity scattering due to the argument by 
Mazin et al. \cite{Mazin}.

\begin{figure}
  \begin{center}
    \includegraphics[width=0.42\columnwidth,angle=270]{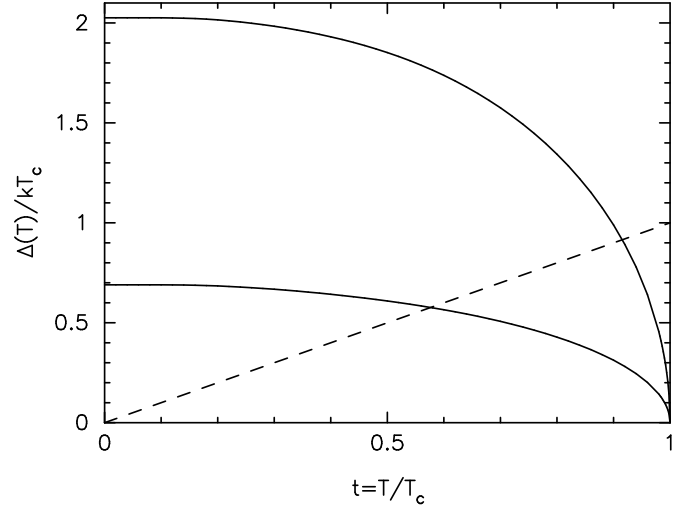}
    \vspace{.1cm}

    \caption{Temperature dependence of the two gaps in MgB$_2$ found from a 
solution of the two by two gap equation. The dashed line shows $\Delta=k_B T$. 
Its crossing point with the small gap appears at a much lower 
temperature than the one with the large gap.
    \label{twogapstemp}}
  \end{center}
\end{figure}

If the intraband scattering rates in the two bands are sufficiently larger 
than their gaps, we can use the Mattis-Bardeen dirty limit formula for the 
conductivity in each band separately \cite{MattisBardeen,SBNam}:
\begin{equation}
\frac{\sigma_\alpha(\omega)}{\sigma_{n,\alpha}} =
\frac{1}{2 \omega} \int_{-\infty}^\infty d \Omega 
\left( \tanh \frac{\Omega + \omega}{2T}- \tanh \frac{\Omega}{2T} \right)
\left[ N_\alpha \left( \Omega \right) 
N_\alpha \left( \Omega + \omega \right) +
M_\alpha \left( \Omega \right) 
M_\alpha \left( \Omega + \omega \right) \right]
\label{MBcond}
\end{equation}
Here, the normal and anomalous densities of states are given by
\begin{equation}
N_\alpha \left( \Omega \right) = {\mathrm{Re}} \left\{
\frac{\left| \Omega \right|}{\sqrt{\Omega^2 - \Delta^2_\alpha}} \right\}
\qquad {\mathrm{and}} \qquad
M_\alpha \left( \Omega \right) = {\mathrm{Re}} \left\{
\frac{\Delta_\alpha \; {\mathrm{sgn}} \left(\Omega \right)}
{\sqrt{\Omega^2 - \Delta^2_\alpha}} \right\}
\end{equation}
The total conductivity under these assumptions is given by $\sigma_1(\omega)=
\sigma_\pi(\omega)+\sigma_\sigma(\omega)$. The temperature dependence of the 
two gaps $\Delta_\alpha$ has to be determined from a solution of the two by 
two gap equation:
\begin{equation}
\Delta_\alpha = \sum_\beta \lambda^{\alpha \beta} \Delta_\beta
\int_0^{\omega_c} dE 
\frac{\tanh \frac{\sqrt{E^2+\Delta_\beta^2}}{2T}}{\sqrt{E^2+\Delta_\beta^2}}
\label{gapeqn}
\end{equation}
Here, $\omega_c$ is a characteristic phonon cut-off frequency and 
$\lambda^{\alpha \beta}$ the two by two coupling matrix. These parameters can 
be either taken from a band structure calculation or tried to be adjusted to 
the particular sample in question. In Ref. \cite{JinPRL} an intermediate approach was 
taken: the partial densities of states and the cut-off frequency were taken 
from band structure calculations, while the remaining parameters were adjusted 
to the $T_c$ of the sample and the value of the small gap, which could be 
extracted from the exponential fall-off of the temperature dependent change 
of the penetration depth. A typical temperature dependence of the two gaps 
found this way is shown in Fig.~\ref{twogapstemp}. One notices that the 
small gap reaches 
$k_B T$ at a much lower temperature than the large gap. According to the rule 
of thumb given above this would mean that the coherence peak in the $\pi$ 
band is expected to appear at much lower temperature than in a conventional 
superconductor.

\begin{figure}
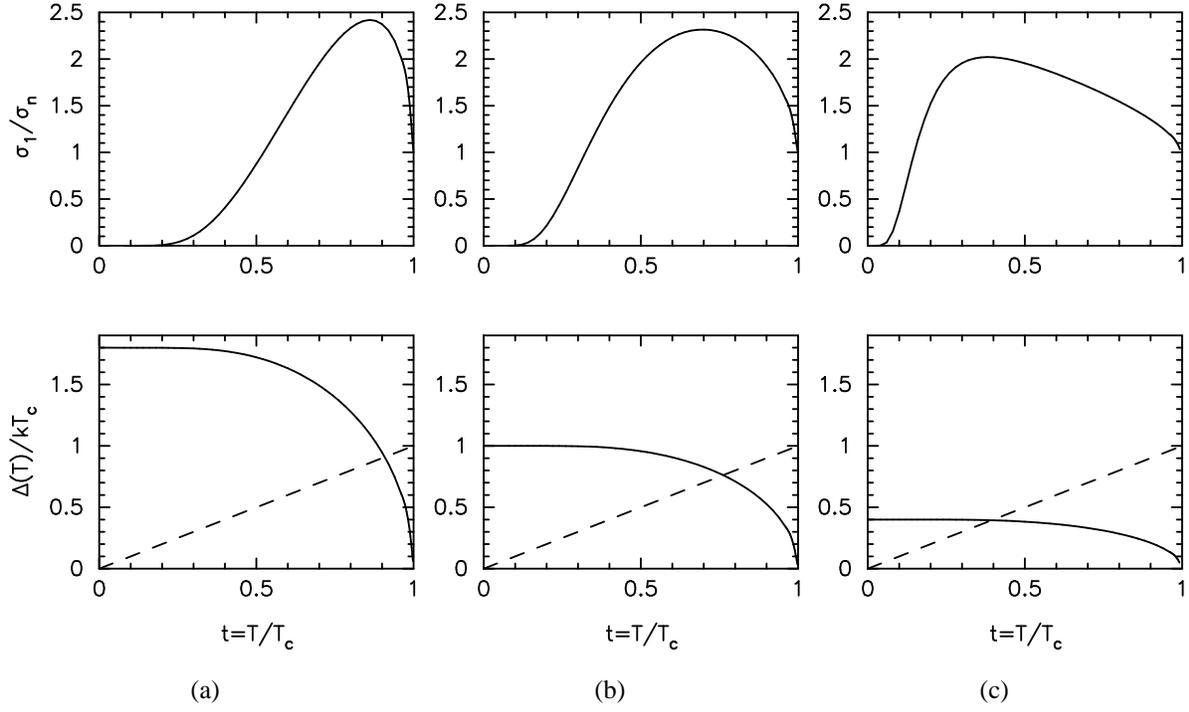

  \begin{center}
\subfigure[]{
    \includegraphics[width=0.54\columnwidth,angle=270]{twoplots18b.ps}}
\subfigure[]{
    \includegraphics[width=0.54\columnwidth,angle=270]{twoplots10b.ps}}
\subfigure[]{
    \includegraphics[width=0.54\columnwidth,angle=270]{twoplots04b.ps}}
    \vspace{.1cm}

    \caption{ Temperature dependence of the microwave conductivity 
(upper panel) for different gap values (lower panel) at a 
frequency of $\omega=0.02 k_B T_c$.
    \label{conddiffgaps}}
  \end{center}
\end{figure}

That this is indeed the case is shown in Fig.~\ref{conddiffgaps}. Here, the 
temperature 
dependence of the Mattis-Bardeen conductivity is shown for different zero 
temperature gap values. In these calculations the BCS temperature dependence 
has been scaled to different gap values for simplicity. Clearly, the position 
of the coherence peak follows the crossing point of the gap with $k_B T$.

In Fig.~\ref{condpeakmgb2} the fits to the two MgB$_2$ samples show the 
temperature dependence 
of the $\pi$ band contribution to the conductivity obtained using the 
temperature dependence of the small gap in Fig.~\ref{twogapstemp}. 
Apparently there is
good agreement between measurement and calculation. The fit to the Niobium 
sample was obtained using the BCS temperature dependence of the gap,
as one should expect for a conventional superconductor.

\begin{figure}
  \begin{center}
    \includegraphics[width=0.42\columnwidth,angle=270]{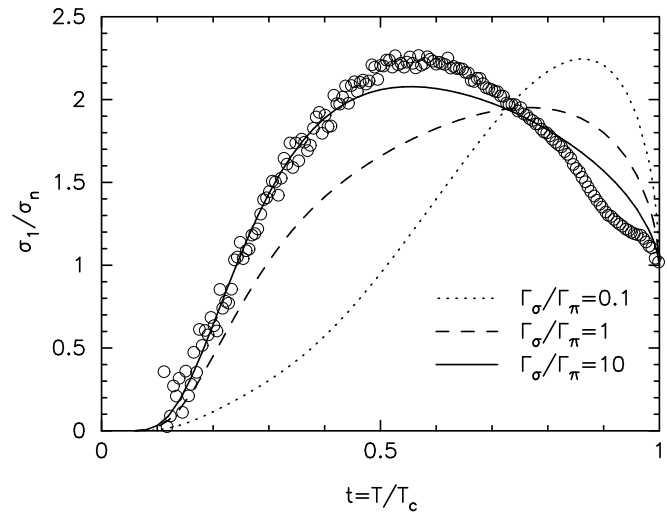}
    \vspace{.1cm}

    \caption{ Temperature dependence of the microwave conductivity for different
ratios of the scattering rates $\Gamma_\sigma/\Gamma_\pi$=0.1, 1, and 10 along
with the experimental results from Ref. \cite{JinPRL}.
    \label{condpeakdiffrat}}
  \end{center}
\end{figure} 

Now one might ask where the contribution from the $\sigma$ band in
the MgB$_2$ samples is. According to Fig.~\ref{conddiffgaps} the
$\sigma$ band should produce a conventional coherence peak. The relative
weight of the two contributions is determined by the partial normal
state conductivities of the two bands, however. We have
\begin{equation}
\sigma_1(\omega)=
\sigma_\pi(\omega)+\sigma_\sigma(\omega) = 
\sigma_{n,\pi} \frac{\sigma_\pi(\omega)}{\sigma_{n,\pi}} +
\sigma_{n,\sigma} \frac{\sigma_\sigma(\omega)}{\sigma_{n,\sigma}} 
\label{relcond}
\end{equation}
where the conductivity ratios are given by Eq.~(\ref{MBcond}). The normal
state conductivities are related to the plasma frequencies
$\omega_{p,\alpha}$ and the intraband scattering rates $\Gamma_\alpha$ via the
expression
\begin{equation} 
\sigma_{n,\alpha} = \hbar \epsilon_0 \frac{\omega^2_{p,\alpha}}{2 \Gamma_\alpha}
\label{signormal}
\end{equation}
Theoretical band structure values for the plasma frequencies have been given
in Ref.~\cite{Mazin}. According to these calculations we have
$\omega_{p,\pi}=5.89$~eV and $\omega_{p,\sigma}=4.14$~eV for the
in plane plasma frequencies. Therefore, if the scattering
rates in the two bands are the same, the contribution of the $\pi$ band
is expected to be a factor of 2 larger than the one of the $\sigma$
band. This is not sufficient to understand the apparent absence of
a sizeable $\sigma$ band contribution in the experimental data. 
A possible explanation could be a stronger scattering rate in the
$\sigma$ band at least in the thin film samples studied in Ref.~\cite{JinPRL}.
To illustrate that, in Fig.~\ref{condpeakdiffrat} the temperature dependence 
of the
conductivity is shown for different values of the relative ratio of 
the two scattering rates. These results suggest a at least 10 times larger
scattering rate in the $\sigma$ band than in the $\pi$ band.
In principle, the scattering rates in the two bands can be varied by
selective doping, for example by doping with Aluminum for Magnesium
or Carbon for Boron \cite{Braccini,Putti,Mazin}. It would be interesting 
to see whether
a cross-over like the one shown in Fig.~\ref{condpeakdiffrat} can be
observed experimentally on a series of MgB$_2$ films with varying degree
of doping.

Concluding this section we have seen that the anomalous microwave conductivity 
peak observed in MgB$_2$ thin films can be understood as a coherence peak
due to the small gap. The peak appears at much lower temperatures than in
conventional superconductors, because the small gap opens up more slowly
upon reducing the temperature below $T_c$. As a result the condensation
of excited quasiparticles and with it the exponential suppression of
the conductivity sets in later. Again, this peculiar effect is consistent
with two gap behavior and reinforces the picture presented in the previous
sections.
\vspace{2ex}

\noindent
{\bf 6. SUMMARY}
\vspace{1ex}

We have reviewed our present picture of superconductivity in magnesium
diboride. Much of this picture stems from band structure calculations.
We have seen that the high value of the critical temperature $T_c$
is due to a high frequency phonon mode that couples strongly to the
electrons at the Fermi level. This coupling is particularly strong
to the electrons on the $\sigma$ band and leads to a much larger
superconducting gap on the $\sigma$ band than on the $\pi$ band.
The different parity of the electronic wavefunctions of the
$\sigma$ band and the $\pi$ band results in a strong reduction
of the interband impurity scattering matrix element. This makes
the two different gaps particularly stable against impurity 
scattering. For these reasons magnesium
diboride appears to be the clearest example of an intrinsic
two gap superconductor to date.

Two gap superconductivity in magnesium diboride leads to 
unusual behavior in a number of experimentally accessible
quantities. Here, we have focused on the temperature dependence
of the upper critical field anisotropy and the microwave
conductivity. The upper critical field anisotropy shows an
unusual strong temperature dependence. This feature is
related to the presence of the two gaps, but also to the
very different anisotropy of the two bands. Here, the
intermediate strength of the interband pairing interaction
plays a crucial role as well.
The temperature dependence of the microwave conductivity
shows a peak at temperatures around 0.5$T_c$. This peak
can be understood as an anomalous coherence peak that
is shifted downwards in temperatures because of the
small gap.

In conclusion, the case of two gap superconductivity in
magnesium diboride generated by strong electron-phonon
coupling appears to be well established and can be
consistently observed in very different experimental
quantities such as the upper critical field or the
microwave conductivity.

The author would like to thank O.~V.~Dolgov, S.~Graser, C.~Iniotakis, 
B.~B.~Jin, N.~Klein, S.-I.~Lee, K.~Maki, A.~I.~Posazhennikova, and 
N.~Schopohl for valuable discussions about this and related topics.
\vspace{2ex}

\newpage
\noindent
{\bf APPENDIX}
\vspace{1ex}

In this appendix the variational method for the calculation of the
upper critical field from Eqs.~(\ref{Eilenbergereq}) and (\ref{gapeqnbc2}) 
shall be described. This method was used in Refs. \cite{Posazh,DahmSchopohl}.

Defining the operator
\begin{equation}
L_\alpha = 2 \left| \omega_n \right | + \; {\rm sgn} \; \omega_n \; 
\vec{v}_{F,\alpha} \left(\hat{k} \right) \left[
\hbar \vec{\nabla} - i \frac{2e}{c} \vec{A} \left( \vec{r} \right)
\right] 
\label{Lop} 
\end{equation}
where $\vec{v}_{F,\alpha}(\hat{k})$ is the Fermi velocity of band $\alpha$,
$\vec{A}$ the vector potential due to the internal
magnetic field within the system, and $\omega_n=(2 n +1) \pi T$ the
Matsubara frequencies,
Eq.~(\ref{Eilenbergereq}) can be inverted using the identity
\begin{equation}
L_{\alpha}^{-1} = \int_0^\infty ds \; e^{-s L_{\alpha}} 
\label{Lopinverse}
\end{equation}
which leads to 
\begin{equation}
f_\alpha (\vec{r}, \hat{k}; \omega_n) = - 2
\int_0^\infty ds \; e^{-s L_{\alpha}} 
\Delta_\alpha (\vec{r})
\end{equation}
Introducing this into the gap equation Eq.~(\ref{gapeqnbc2})
we can eliminate $f_\alpha$:
\begin{equation}
\Delta_\alpha (\vec{r}) = 2 \pi T
\sum_{\alpha'} \sum_{|\omega_n| < \omega_c} \lambda^{\alpha \alpha'} 
\left\langle
\int_0^\infty ds \; e^{-s L_{\alpha'}} 
\Delta_{\alpha'} (\vec{r})
\right\rangle_{\alpha'}
\label{gapeqnbc2delta}
\end{equation}
This equation is a linear equation for $\Delta_\alpha (\vec{r})$.

As is usual in weak-coupling theory we can eliminate the
cutoff frequency $\omega_c$ 
in favor of the critical temperature $T_c$. For this purpose
we consider Eq.~(\ref{gapeqnbc2delta}) in the absence of
a magnetic field at $T_c$. Then the gap function 
$\Delta_\alpha (\vec{r})$ becomes homogeneous and we find 
\begin{eqnarray}
\Delta_\alpha &=& 2 \pi T_c
\sum_{|\omega_n \left( T_c \right)| < \omega_c} 
\int_0^\infty ds \; e^{-2 s \left| \omega_n \left( T_c \right) \right |} 
\sum_{\alpha'} \lambda^{\alpha \alpha'} \Delta_{\alpha'} \nonumber \\
 &=& 2 \pi T_c
\sum_{\omega_n \left( T_c \right) > 0}^{\omega_c}
\frac{1}{\omega_n \left( T_c \right)} 
\sum_{\alpha'} \lambda^{\alpha \alpha'} \Delta_{\alpha'} 
\end{eqnarray}
This is an eigenvalue equation for $\Delta_\alpha$ and the largest
eigenvalue $\lambda_+$ of the matrix $\lambda^{\alpha \alpha'}$
determines $T_c$. Thus we find
\begin{equation}
\frac{1}{\lambda_+} = 2 \pi T_c
\sum_{\omega_n \left( T_c \right) > 0}^{\omega_c}
\frac{1}{\omega_n \left( T_c \right)}
\label{invlambdaplus}
\end{equation}
For $\omega_c \gg T_c$ the following relation holds:
\begin{equation}
2 \pi T_c 
\sum_{\omega_n(T_c) > 0}^{\omega_c} \frac{1}{\omega_n(T_c)} \; - \; 
2 \pi T
\sum_{\omega_n(T) > 0}^{\omega_c} \frac{1}{\omega_n(T)}
\simeq \ln \frac{T}{T_c}
\label{eq9}
\end{equation}
and thus we can write Eq.~(\ref{invlambdaplus})
in the form
\begin{equation}
\frac{1}{\lambda_+} - \ln \frac{T}{T_c} = 2 \pi T
\sum_{\omega_n(T) > 0}^{\omega_c}
\frac{1}{\omega_n \left( T \right)} = 2 \pi T 
\sum_{|\omega_n(T)| < \omega_c} 
\int_0^\infty ds e^{- 2 s |\omega_n(T)|}
\end{equation}
Multiplying this equation by 
$\sum_{\alpha'} \lambda^{\alpha \alpha'} 
\Delta_{\alpha'} (\vec{r})$ it can be subtracted from
Eq.~(\ref{gapeqnbc2delta}) leading to
\begin{eqnarray}
\lefteqn{
\Delta_\alpha (\vec{r}) + \sum_{\alpha'} \lambda^{\alpha \alpha'} 
\Delta_{\alpha'} (\vec{r}) \left( - \frac{1}{\lambda_+} + 
\ln \frac{T}{T_c} \right) }
\nonumber \\ &=& 
\sum_{\alpha'} \lambda^{\alpha \alpha'} 
2 \pi T \sum_{|\omega_n| < \omega_c} \left\langle
\int_0^\infty ds \; \left[ e^{-s L_{\alpha'}} - 
e^{- 2 s |\omega_n|} \right] \Delta_{\alpha'} (\vec{r}) 
\right\rangle_{\alpha'}  \nonumber \\
&=& 
\sum_{\alpha'} \lambda^{\alpha \alpha'} 
4 \pi T \sum_{\omega_n > 0}^{\omega_c}
\int_0^\infty ds \; e^{- 2 s \omega_n} \left\langle 
\left[ e^{-i s \vec{v}_{F,\alpha'} \cdot \vec{\Pi}} - 1
\right] \Delta_{\alpha'} (\vec{r}) 
\right\rangle_{\alpha'} 
\label{eqa11}
\end{eqnarray}
Here, we have eliminated the sgn~$\omega_n$ factor
assuming inversion symmetry of the Fermi velocity
$\vec{v}_{F,\alpha} \left(\hat{k} \right)=
-\vec{v}_{F,\alpha} \left(-\hat{k} \right)$ and
introduced the operator
\begin{equation}
\vec{\Pi} = \frac{\hbar}{i} \vec{\nabla} - \frac{2e}{c} \vec{A} \left( \vec{r} \right)
\end{equation}
In Eq. (\ref{eqa11}) we may now extend the $\omega_n$ summation to
infinity and sum it up:
\begin{eqnarray}
\sum_{\omega_n > 0}
e^{- 2 s \omega_n} & = &  \sum_{n=0}^\infty e^{- 2 s (2 n + 1) \pi T}
=  e^{- 2 s \pi T} \sum_{n=0}^\infty \left( e^{- 4 s \pi T} \right)^n
\nonumber \\
& = &  e^{- 2 s \pi T} \frac{1}{1 - e^{- 4 s \pi T} }
= \frac{1}{2 \sinh \left( 2 s \pi T \right)}
\label{eqa13}
\end{eqnarray}
Note that integration and summation in Eq.~(\ref{eqa11}) may only be 
interchanged because the divergence for $s \rightarrow 0$ has been
eliminated. (In Eq.~(\ref{gapeqnbc2delta}) this summation was not possible).
So we find from Eq.~(\ref{eqa11}):
\begin{eqnarray}
\lefteqn{
\Delta_\alpha (\vec{r}) + \sum_{\alpha'} \lambda^{\alpha \alpha'} 
\Delta_{\alpha'} (\vec{r}) \left( - \frac{1}{\lambda_+} + 
\ln \frac{T}{T_c} \right) }
\nonumber \\ &=& 
\sum_{\alpha'} \lambda^{\alpha \alpha'} 
\int_0^\infty \frac{du}{\sinh u}
\left\langle 
\left[ e^{-i u \vec{v}_{F,\alpha'} \cdot \vec{\Pi} / 
\left( 2 \pi T \right)} - 1
\right] \Delta_{\alpha'} (\vec{r}) 
\right\rangle_{\alpha'} 
\label{eqa14}
\end{eqnarray}
In the presence of an external magnetic field $\vec{B} = \vec{\nabla} \times
\vec{A}$ it is convenient to choose the field direction as the
$z$-axis of the coordinate system. In these coordinates $\Delta_\alpha (\vec{r})$
does not depend on $z$. Choosing the gauge $\vec{A} = B x \hat{y}$ the operator
$\vec{\Pi}$ simplifies to
\begin{equation}
\vec{\Pi} = \frac{\hbar}{i} \vec{\nabla} - \frac{2e}{c} \vec{A} \left( \vec{r} \right)
= \left( \begin{array}{c} -i\hbar {\partial}_x \\  -i\hbar {\partial}_y
- \frac{2e}{c} B x \\ 0 \end{array} \right) = \sqrt{\frac{e B}{c}}
\left( \begin{array}{c} a + a^\dagger \\  i (a - a^\dagger)
\\ 0 \end{array} \right)
\end{equation}
where we have introduced raising and lowering operators
$2 \sqrt{\frac{e B}{c}} a =  -i \hbar {\partial}_x - \hbar {\partial}_y + 2 i \frac{e B}{c} x$ and
$2 \sqrt{\frac{e B}{c}} a^\dagger =  -i \hbar {\partial}_x + \hbar {\partial}_y - 2 i \frac{e B}{c} x$.
Thus, we see from Eq. (\ref{eqa14}) that only the components of the
Fermi velocity perpendicular to the magnetic field direction play
a role and we have
\begin{equation}
\vec{v}_{F,\alpha} \cdot \vec{\Pi} = \sqrt{\frac{e B}{c}} \left[ 
\left( v_{x,\alpha} + i v_{y,\alpha} \right) a + 
\left(  v_{x,\alpha} - i v_{y,\alpha} \right) a^\dagger \right]
\label{eqa16}
\end{equation}
where $v_{x,\alpha}$ and $v_{y,\alpha}$ are the components of the Fermi velocity
perpendicular to the magnetic field $\vec{B}$. Note, that these components
are not necessarily identical to the components of the Fermi velocity with
respect to the crystal axes in Eq.~(\ref{Fermivel}).

It is useful to introduce a scaling
of the $x$- and $y$-coordinates of the form
\begin{equation}
x = e^{\tau} \bar{x} \qquad y = e^{-\tau} \bar{y}
\label{eq17}
\end{equation}
Here $e^{\tau}$ is a scaling factor which scales the $x$- and 
$y$-coordinates in such a way as to preserve the area.
Expressed in the new coordinates $\bar{x}$ and $\bar{y}$
the operator $\vec{\Pi}$ can be written:
\begin{equation}
\vec{\Pi}
= \left( \begin{array}{c} -i\hbar e^{-\tau} {\partial}_{\bar{x}} \\  
-i\hbar e^{\tau} {\partial}_{\bar{y}}
- \frac{2e}{c} e^{\tau} B \bar{x} \\ 0 \end{array} \right) = \sqrt{\frac{e B}{c}}
\left( \begin{array}{c} e^{-\tau} (\bar{a} + \bar{a}^\dagger) \\  
i e^{\tau} (\bar{a} - \bar{a}^\dagger)
\\ 0 \end{array} \right)
\end{equation}
where the raising and lowering operators $\bar{a}$ and $\bar{a}^\dagger$
are also expressed in terms of these new coordinates. Using this
result we find
\begin{equation}
\vec{v}_{F,\alpha} \cdot \vec{\Pi} = \sqrt{\frac{e B}{c}} \left[ 
\left( e^{-\tau} v_{x,\alpha} + i e^{\tau} v_{y,\alpha} \right) \bar{a} + 
\left( e^{-\tau} v_{x,\alpha} - i e^{\tau} v_{y,\alpha} \right) \bar{a}^\dagger \right]
\end{equation}

We can now write the exponential operator in Eq. (\ref{eqa14}) as follows:
\begin{equation}
e^{- i u \vec{v}_{F,\alpha} \cdot \vec{\Pi} / (2 \pi T) } =
e^{A + B}
\label{eqa20}
\end{equation}
with
\begin{eqnarray}
A & = & - i \frac{u}{2 \pi T} \sqrt{\frac{e B}{c}}
\left( e^{-\tau} v_{x,\alpha} - i e^{\tau} v_{y,\alpha} \right) \bar{a}^\dagger 
\nonumber \\
B & = &  - i \frac{u}{2 \pi T} \sqrt{\frac{e B}{c}} 
\left( e^{-\tau} v_{x,\alpha} + i e^{\tau} v_{y,\alpha} \right) \bar{a}
\label{eqa21}
\end{eqnarray}
Using the identity
\begin{equation}
e^{A + B} = e^{-\left[ A,B \right]/2} \; e^A \; e^B
\end{equation}
and calculating the commutator
\begin{equation}
\left[ A,B \right] = u^2 \frac{e B}{c \left( 2 \pi T \right)^2}
\left(  e^{-2\tau} v_{x,\alpha}^2 + e^{2\tau} v_{y,\alpha}^2 \right) 
\end{equation}
we can write
\begin{eqnarray}
\lefteqn{
e^{- i u \vec{v}_{F,\alpha} \cdot \vec{\Pi} / (2 \pi T) } =} \nonumber \\ & &
e^{-u^2 \frac{e B}{2 c \left( 2 \pi T \right)^2} 
\left( e^{-2\tau} v_{x,\alpha}^2 + e^{2\tau} v_{y,\alpha}^2 \right) } \; 
e^{- i \frac{u}{2 \pi T} \sqrt{\frac{e B}{c}}
\left( e^{-\tau} v_{x,\alpha} - i e^{\tau} v_{y,\alpha} \right) \bar{a}^\dagger } \; 
e^{- i \frac{u}{2 \pi T} \sqrt{\frac{e B}{c}}
\left( e^{-\tau} v_{x,\alpha} + i e^{\tau} v_{y,\alpha} \right) \bar{a} }
\nonumber
\end{eqnarray}

In principle, we can view Eq. (\ref{eqa14}) as an eigenvalue
problem. The highest eigenvalue and its corresponding eigenfunction
of the right hand side operator will determine the upper critical
field $B_{c2}(T)$. We already know from Abrikosov's solution of
Ginzburg-Landau theory
that the solution for an isotropic $s$-wave superconductor will be
Abrikosov's vortex state wave function $\psi_\Lambda (\vec{r})$. 
This wavefunction has the property
that it is destroyed by the operator $a$
\begin{equation}
a \; \psi_\Lambda (\vec{r}) = 0
\end{equation}
and has the form of the lowest Landau level wavefunction. In principle,
one can try to solve Eq. (\ref{eqa14}) by making a Landau level expansion
of $\Delta_\alpha (\vec{r})$ \cite{Rieck,SunMaki}. However, for an anisotropic 
superconductor one
should expect a distortion of the vortex lattice. Therefore here we will
start from a variational ansatz to Eq. (\ref{eqa14}) by choosing a
different wavefunction, which obeys the equation
\begin{equation}
\bar{a} \; \Delta_\alpha (\vec{r}) = 0
\end{equation}
This corresponds to the choice Eq.~(\ref{varground}).
Here, we can use the scaling parameter $\tau$ now as a variational
parameter which has to be adjusted such as to yield the highest
possible eigenvalue of Eq.~(\ref{eqa14}). If we insert this
variational wavefunction into Eq.~(\ref{eqa14}) the $e^B$ and $e^A$
operators just drop out and we are left with the
equation
\begin{equation}
\Delta_\alpha = 
\sum_{\alpha'} \lambda^{\alpha \alpha'} \left[ \frac{1}{\lambda_+} -
\ln \frac{T}{T_c} - l_{\alpha'} (\tau, \frac{B_{c2}}{T^2} ) \right] \Delta_{\alpha'}
\label{eqds3}
\end{equation}
where the function $l_{\alpha}$ is given by the expression
\begin{equation}
l_{\alpha} \left( \tau, \frac{B_{c2}}{T^2} \right)=
\int_0^\infty \frac{du}{\sinh u} \left\langle 1 -
e^{-u^2 \frac{eB_{c2}}{8 c \pi^2 T^2} \left( e^{-2\tau} v_{x, \alpha}^2
(\hat{k}) +  e^{2\tau} v_{y, \alpha}^2 (\hat{k}) \right) }
\right\rangle_\alpha \label{eqds4}
\end{equation}
Equation (\ref{eqds3}) is a $2\times 2$ matrix equation and the
upper critical field is determined by the criterion that the
largest eigenvalue of the right hand side becomes 1.
This criterion leads to the characteristic equation
\begin{equation}
(1-\eta) l_\sigma + \eta \; l_\pi + \ln t =
- \Lambda_{\pm}
\left( l_\sigma + \ln t \right)
\left( l_\pi + \ln t \right)
\label{eqchar}
\end{equation}
where $t=T/T_c$ and the parameters $\eta$ and $\Lambda_{\pm}$ are given by
\begin{equation}
\eta = \frac{\lambda^{\pi \pi} - \lambda_-}{\lambda_+ - \lambda_-}
\qquad \mbox{and} \qquad
\Lambda_{\pm}=\frac{\lambda_+ \lambda_-}{\lambda_+ - \lambda_-}
\end{equation}
Here, $\lambda_+$ and $\lambda_-$ are the larger and smaller eigenvalue of 
the matrix $\lambda^{\alpha \alpha'}$, respectively. Note, 
that Eq.~(\ref{eqchar})
means that only the two parameters $\eta$ and $\Lambda_{\pm}$
out of the four parameters $\lambda^{\alpha \alpha'}$ determine
the upper critical field of a two gap superconductor.

Equation (\ref{eqchar}) is a quadratic equation in $\ln t$ and allows to
calculate $t$ once $l_\pi$ and $l_\sigma$ are known. Therefore,
calculation of $B_{c2}$ from Eqs.~(\ref{eqds4}) and (\ref{eqchar})
can proceed as follows: for given values of $B_{c2}/T^2$ and
$\tau$ the quantities $l_\pi$ and $l_\sigma$ are calculated 
using Eq.~(\ref{eqds4}) and the Fermi velocities given in
Eq.~(\ref{Fermivel}).
Then $t$ can be calculated using Eq.~(\ref{eqchar}) and
the matrix elements given in
Eq.~(\ref{lambdamat}). Now, the
parameter $\tau$ is optimized such as to maximize $t$.
From the maximized $t$ and the given value of $B_{c2}/T^2$ 
finally $B_{c2}$ can be calculated. This procedure is
repeated for several values of $B_{c2}/T^2$ producing
a $B_{c2}(T)$ curve.

In Eq.~(\ref{eqds4}) the integration over $u$ has a 
special form and it is useful to introduce the integral
\begin{equation}
I(y)
=\int_0^\infty \frac{du}{\sinh u}
\left( 1 - e^{-y u^2} \right)
\label{eqa28}
\end{equation}
While this integral cannot be calculated analytically in the
most general case, it at least possesses some simple limiting
expressions for small and large arguments $y$. (Large $y$ corresponds
to the zero temperature limit in Eq. (\ref{eqds4}), while small
$y$ corresponds to $T \rightarrow T_c$, when $B_{c2} \rightarrow 0$.)
These limiting expressions read
\begin{equation}
I(y) = \left\{
\begin{array}{c@{\quad {\rm for} \quad}c}
\frac{7}{2} \zeta(3) y - \frac{93}{4} \zeta(5) y^2 & y \ll 1 \\[1ex]
\frac{1}{2} \ln (4 \gamma y) & y \gg 1
\end{array}
\right.
\label{eqa29}
\end{equation}
Here, $\zeta(n)$ is the Riemann Zeta function and $\ln \gamma=0.577215$ 
Euler's constant. Using the integral $I(y)$ Eq. (\ref{eqds4}) can be written
\begin{equation}
l_{\alpha} \left( \tau, \frac{B_{c2}}{T^2} \right)=
\left\langle I \left[
\frac{eB_{c2}}{8 c \pi^2 T^2} \left( e^{-2\tau} v_{x, \alpha}^2
(\hat{k}) +  e^{2\tau} v_{y, \alpha}^2 (\hat{k}) \right) \right]
\right\rangle_\alpha 
\end{equation}
Using Eq.~(\ref{eqa29}) this allows to find simple limiting expressions
for $l_{\alpha}$ in the limits $T \rightarrow 0$ and $T \rightarrow T_c$.
In particular, in the limit $T \rightarrow T_c$ we have in linear order
in $B_{c2}$
\begin{equation}
l_{\alpha} =
\frac{7}{2}\zeta(3) \frac{eB_{c2}}{8 c \pi^2 T_c^2} 
\left( e^{-2\tau} \left\langle v_{x, \alpha}^2 \right\rangle_\alpha 
+  e^{2\tau} \left\langle v_{y, \alpha}^2 \right\rangle_\alpha \right) 
\end{equation}
Since $B_{c2}$ goes to zero for $T \rightarrow T_c$, both $l_{\alpha}$
and $\ln t$ decrease linearly and the right hand side of Eq.~(\ref{eqchar})
can be neglected, because it becomes quadratic in $1-t$. Thus in
linear order we have
\begin{eqnarray}
\lefteqn{- \ln \frac{T}{T_c} = (1-\eta) l_\sigma + \eta \; l_\pi =
\frac{7\zeta(3)eB_{c2}}{16 c \pi^2 T_c^2} \cdot } \nonumber \\
&& \cdot \left\{ e^{-2\tau} \left[ 
(1-\eta) \left\langle v_{x, \sigma}^2 \right\rangle_\sigma +
\eta \left\langle v_{x, \pi}^2 \right\rangle_\pi \right]
+  e^{2\tau} \left[ 
(1-\eta) \left\langle v_{y, \sigma}^2 \right\rangle_\sigma +
\eta \left\langle v_{y, \pi}^2 \right\rangle_\pi \right] \right\} 
\label{bc2tc}
\end{eqnarray}
Minimizing this expression with respect to $\tau$ one finds
\begin{equation}
e^{2\tau} = \sqrt{\frac{(1-\eta) \left\langle v_{x, \sigma}^2 \right\rangle_\sigma +
\eta \left\langle v_{x, \pi}^2 \right\rangle_\pi }
{(1-\eta) \left\langle v_{y, \sigma}^2 \right\rangle_\sigma +
\eta \left\langle v_{y, \pi}^2 \right\rangle_\pi}}
\end{equation}
Using this result we can calculate the slope of $B_{c2}$ at $T_c$ from
Eq.~(\ref{bc2tc}):
\begin{equation}
\left. \frac{d B_{c2}}{d T} \right|_{T_c} = \frac{8 c \pi^2 T_c}{7\zeta(3)e}
\frac{1}{\sqrt{\left[ (1-\eta) \left\langle v_{x, \sigma}^2 \right\rangle_\sigma +
\eta \left\langle v_{x, \pi}^2 \right\rangle_\pi \right]
\left[ (1-\eta) \left\langle v_{y, \sigma}^2 \right\rangle_\sigma +
\eta \left\langle v_{y, \pi}^2 \right\rangle_\pi  \right] } }
\end{equation}

\end{document}